\documentclass{optica-article}

\journal{opticajournal}

\articletype{Research Article}

\usepackage{siunitx}
\usepackage{lineno}

\begin{document}

\title{High-power ultra-flat supercontinuum generation in hydrogen- and deuterium-filled hollow-core antiresonant fiber}

\author{Athanasios Lekosiotis\authormark{1}, Balazs Plosz\authormark{1}, Federico Belli\authormark{1}, Mohammed Sabbah\authormark{1}, Amir Abdolvand\authormark{2} and John C. Travers\authormark{1,*}}

\address{\authormark{1}School of Engineering and Physical Sciences, Heriot-Watt University, Edinburgh, EH14 4AS, UK}
\address{\authormark{2}ASML Research, ASML Netherlands B.V., 5504 DR Veldhoven, Netherlands}

\email{\authormark{*}j.travers@hw.ac.uk}

\begin{abstract*}
We demonstrate Raman-driven supercontinuum generation with high spectral flatness in hydrogen- and deuterium-filled antiresonant hollow-core fibers. Using 400~fs pulses at 515~nm, a vibrational Raman frequency comb is generated and subsequently smoothed into a uniform supercontinuum through self-phase modulation and rotational Raman scattering. Unlike nitrogen and methane, hydrogen enables power scaling to high repetition rates without gain suppression or thermal degradation. Single-pump excitation at 515~nm generates a supercontinuum with a \SI{-10}{\decibel} span (relative to the spectral peak) of 403--1469~nm in H$_2$ and 385--1295~nm in D$_2$, both extending to $\sim$350~nm at the \SI{-20}{\decibel} level. At 2.5~MHz repetition rate (25~W average input power), the spectral power density exceeds 5~mW/nm over 491--869~nm and 10~mW/nm over 509--769~nm, corresponding to approximately 5~W of total output average power. When using a dual-pumping configuration at both 515~nm and 1030~nm, the supercontinuum extends to around 1650~nm at the \SI{-20}{\decibel} level with improved conversion efficiency. These results establish hydrogen- and deuterium-filled hollow-core fibers as a platform for high-power broadband spatially coherent light sources.
\end{abstract*}

\section{Introduction}\label{sec:intro}
Supercontinuum generation in gas-filled hollow-core antiresonant fibers provides an excellent route towards broadband spatially coherent light sources with high brightness~\cite{travers_ultrafast_2011, russell_hollow-core_2014}. Compared to solid-core silica fibers, the use of a gaseous nonlinear medium enables spectral extension deeper into the ultraviolet, avoiding the photodarkening and solarization effects that limit silica-based sources~\cite{travers_blue_2010, sylvestre_recent_2021}. Furthermore, the high damage threshold and broad transparency of gas media have enabled vacuum-ultraviolet supercontinuum generation~\cite{belli_vacuum-ultraviolet_2015, ermolov_supercontinuum_2015} and multi-octave continua spanning from the deep ultraviolet to the mid-infrared~\cite{elu_seven-octave_2021, hosseini_uv_2018, adamu_deep-uv_2019}.

Commercial Yb-based laser systems now provide hundreds of watts of average power at high repetition rates, making them attractive pump sources for high-power supercontinuum generation. However, the relatively long pulse durations typical of such systems (a few hundred femtoseconds) are not well suited to established supercontinuum generation techniques. Soliton self-compression~\cite{travers_high-energy_2019} and resonant dispersive wave emission~\cite{joly_bright_2011} typically require few-cycle or sub-30~fs pump pulses, while self-phase modulation in the normal dispersion regime provides only limited spectral broadening for longer pulses. Modulational instability can generate broadband spectra with longer pulses~\cite{tani_phz-wide_2013}, but the resulting supercontinuum exhibits high relative intensity noise and low temporal coherence due to its noise-seeded nature~\cite{Dudley_2006}.

An alternative approach that overcomes these limitations is stimulated Raman scattering (SRS) in molecular gases~\cite{mousavi_nonlinear_2018, Gladyshev_2019, Gladyshev2022}. We recently demonstrated~\cite{gao_raman_2022} a technique we term comb-to-continuum (CtC), which proceeds in two stages: (i) vibrational Raman scattering generates a broad frequency comb spanning from the ultraviolet to the infrared; (ii) the discrete comb lines are subsequently broadened and merged into a smooth supercontinuum through self-phase modulation and rotational Raman scattering. Each comb line corresponds, in the time domain, to a short transient embedded within the overall pump envelope; because the local intensity varies across the envelope, neighboring transients accumulate different nonlinear phase shifts, and this distribution of phases broadens each individual comb tooth rather than producing further cascaded lines~\cite{gao_raman_2022}. The non-instantaneous rotational Raman response contributes to the same line-filling step. This technique offers several key advantages. The spectral extent of the supercontinuum is established by the vibrational Raman comb, which can span multiple octaves even from a narrowband pump. Unlike spectral broadening based on self-phase modulation or soliton dynamics, which proceed continuously from the pump wavelength and can be blocked or reduced by high-loss fiber resonance bands, the large frequency spacing of the Raman comb allows it to jump over such bands entirely. Additionally, the two stages of the process respond differently to changes in gas pressure: the vibrational Raman frequency shift is an intrinsic molecular constant, whereas the Kerr nonlinearity that drives the line broadening scales with gas density, providing an extra degree of control over the resulting spectrum. The process works efficiently with relatively long pump pulses (hundreds of femtoseconds to tens of picoseconds), eliminating the need for complex ultrashort pulse sources. Finally, pumping in the normal dispersion regime maximizes vibrational Raman gain while suppressing modulational instability.

The original comb-to-continuum demonstration~\cite{gao_raman_2022} used nitrogen as the filling gas, pumped at 532~nm with 20~ps pulses, producing a supercontinuum spanning 440--1200~nm. However, subsequent investigations revealed that nitrogen's closely spaced rotational Raman lines cause gain suppression by coupling energy into higher-order fiber modes~\cite{sabbah_effect_2023}, limiting both the spectral extent and flatness of the supercontinuum, and degrading the mode quality at the output. Methane, which has no rotational Raman response due to its symmetric molecular structure, was explored as an alternative and delivered broadband spectra spanning 350--1700~nm~\cite{plosz_supercontinuum_2025}. However, linear and nonlinear absorption in methane leads to thermal decomposition of the gas at moderate repetition rates, causing fiber damage and preventing power scaling beyond approximately 50~kHz.

Hydrogen and deuterium offer the ideal combination of properties for high-power Raman-driven supercontinuum generation. Their strong vibrational Raman responses, with large frequency shifts of approximately 125~THz and 90~THz respectively, enable efficient broadband comb generation. Critically, their rotational Raman lines are widely spaced, avoiding the gain suppression observed in nitrogen. They also possess high transparency from the deep ultraviolet to the infrared and very high thermal conductivity, preventing the thermal issues that limit methane. These properties make hydrogen and deuterium uniquely suited for scaling the CtC technique to high average powers. Hydrogen- and deuterium-filled hollow-core fibers have a long history as a platform for narrowband Raman frequency converters and fiber-gas lasers, from the first low-threshold SRS demonstrations~\cite{benabid_stimulated_2002, couny_subwatt_2007} to high-power continuous-wave and pulsed Raman lasers operating into the mid-infrared~\cite{Cui:23}. Here we exploit the same favorable Raman and thermal properties to drive broadband supercontinuum generation at high average power.

Scaling gas-filled hollow-core fiber sources to high average power is essential for practical applications but presents distinct challenges. Although the fiber waveguide itself can handle very high average powers---kilowatts of continuous-wave light have been transmitted through air-filled antiresonant fibers~\cite{mulvad_kilowatt-average-power_2022, cooper_single-mode_2023}---the gas medium introduces its own constraints at high repetition rates. Energy deposited through photoionization and/or Raman scattering heats the gas on timescales of nanoseconds to microseconds, perturbing the refractive index profile within the core and thereby affecting the propagation of subsequent pulses~\cite{koehler_long-lived_2018, tani_long-lived_2025}. At megahertz repetition rates, cumulative heating can build up between pulses before the gas fully recovers~\cite{koehler_post-recombination_2021}. For molecular gases in particular, the elevated temperature redistributes the rotational level populations, which in turn modifies the effective third-order nonlinearity~\cite{beetar_thermal_2021}. These thermal constraints have been quantified for soliton-effect pulse compression~\cite{schade_scaling_2021}, but for Raman-driven supercontinuum generation they have not yet been explored.

In this work, we demonstrate high-power supercontinuum generation with high spectral flatness in hydrogen- and deuterium-filled antiresonant fibers pumped at 515~nm. We denote this single-wavelength pumping configuration as ``single-pumped''. Using 400~fs pulses from a frequency-doubled Yb laser, we generate supercontinua with a \SI{-10}{\decibel} span of 403--1469~nm in H$_2$ and 385--1295~nm in D$_2$, with high flatness even on a linear scale. By increasing the repetition rate to 2.5~MHz (25~W average input power), we generate a supercontinuum with spectral power density exceeding 5~mW/nm over the visible--near-infrared range (491--869~nm) and 10~mW/nm over 509--769~nm, corresponding to approximately 5~W of total output average power, without any sign of saturation. To further enhance the infrared portion of the spectrum and improve overall efficiency, we also demonstrate a ``dual-pumped'' configuration using both 515~nm and 1030~nm wavelengths simultaneously. Dual-wavelength pumping schemes have previously been explored in solid-core fibers as a means of tailoring spectral broadening dynamics~\cite{champert_white-light_2004}. Here, we extend this concept to the CtC scheme in gas-filled antiresonant fibers, enabling an ultra-flat supercontinuum with a \SI{-10}{\decibel} span of 439--1547~nm at reduced 515~nm pulse energies and with significantly improved conversion efficiency.

\section{Experimental setup}\label{sec:setup}
\begin{figure}[tbh]
\centering
\includegraphics[width=5in]{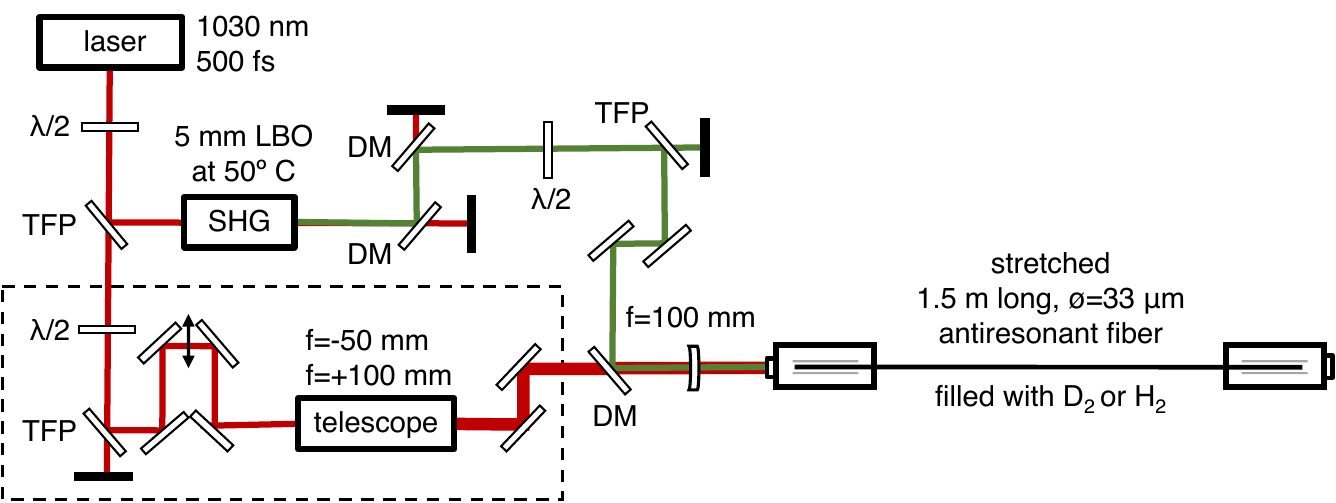}
\caption{Experimental setup for single-pumping (515~nm only, without optics in dashed rectangle) or dual-pumping (515~nm and 1030~nm, with optics in dashed rectangle). Pulses at 515~nm were generated in a 5~mm thick LBO crystal heated to \qty{50}{\celsius} and focused into a 1.5~m long, \SI{33}{\micro\meter} core diameter antiresonant fiber filled with deuterium (D$_2$) or hydrogen (H$_2$). The 1030~nm beam was magnified by a lens telescope for optimal mode-matching at the fiber input. $\lambda$/2: half-wave plate; TFP: thin-film polarizer; DM: dichroic mirror. The vertical arrow indicates a delay stage for temporal synchronization.}
\label{fig:setup}
\end{figure}

A sketch of the experimental layout is shown in Fig.~\ref{fig:setup}. The output of a commercial Yb-based laser (1030~nm center wavelength, 500~fs pulse duration) was split using a half-wave plate ($\lambda$/2) and a thin-film polarizer (TFP) to create two pump lines: one at the fundamental wavelength (1030~nm) and another at the second harmonic (515~nm). For single-pumped operation, only the 515~nm beam was used; for dual-pumped operation, both wavelengths were combined at the fiber input.

Second harmonic generation (SHG) was performed in a 5~mm thick lithium triborate (LBO) crystal heated to \qty{50}{\celsius} using a home-built oven. Residual 1030~nm light was removed using two dichroic mirrors. A half-wave plate paired with a thin-film polarizer provided independent power control for the 515~nm beam. The beam was coupled into the fiber using a fused silica lens with f~=~100~mm focal length, mounted on a 3-axis translation stage.

The 1030~nm beam transmitted through the first polarizer passed through a motorized delay stage for temporal synchronization with the 515~nm pulses. The beam was magnified using a lens telescope consisting of a diverging lens (f~=~$-$50~mm) and a converging lens (f~=~+100~mm), with adjustable separation to control the collimation. Two steering mirrors before the dichroic mirror (DM) allowed optimization of the 1030~nm beam pointing. The DM combined the 1030~nm beam with the reflected 515~nm beam. The telescope lens separation was adjusted to match the 1030~nm focal spot size and position to that of the 515~nm beam, enabling both wavelengths to be coupled into the fiber using the same focusing lens.

Two single-ring antiresonant hollow-core fibers were used in this work. Both had the same overall geometry: a 1.5~m length, a \SI{33}{\micro\meter} core diameter, and a cladding consisting of six non-touching glass resonators surrounding the central hollow core. They differed only in the thickness of the core-wall (resonator) silica membranes. The first fiber (hereafter ``thin-wall fiber'') had a wall thickness of approximately 160~nm, placing its first ($m=1$) antiresonant cladding resonance at $\sim$348~nm. The second fiber (hereafter ``thick-wall fiber'') had a wall thickness of approximately 174~nm, shifting the resonance to $\sim$376~nm and thereby restricting the short-wavelength edge of the supercontinuum while reducing loss in the near-infrared. The thin-wall fiber was used for all single-pumped results reported in Section~\ref{sec:SP} (Figs.~\ref{fig:escans}, \ref{fig:515nmpumping} and \ref{fig:SPSC_PS}), while the thick-wall fiber was used for the dual-pumping experiments of Section~\ref{sec:DP} (Figs.~\ref{fig:DP} and \ref{fig:DP_PS}).

The fiber resonators were optimized to provide single-mode performance around 1030~nm~\cite{uebelBroadbandRobustlySinglemode2016}, and the thin walls designed to provide broadband guidance across the visible and near-infrared. However, in this work we pumped at 515~nm, which, without special care, can excite higher-order modes at the input or cause inter-mode coupling along the fiber. To prevent this, we stretched and mounted the fiber in the same way as for large-core hollow capillary fibers~\cite{nagyFlexibleHollowFiber2008}, which maintained excellent single-mode transmission at 515~nm and across the supercontinuum.

The fiber was mounted in a gas-handling system consisting of two cells connected by a steel tube, with ports for gas filling and evacuation. Optical access was provided by 3~mm thick uncoated fused silica windows. The window thickness was chosen to minimize both mechanical stress from the gas pressure and Kerr lensing from the high-energy input pulses~\cite{brahms_effect_2023}.

Output spectra were fully captured using an integrating sphere and measured using a pair of Avantes spectrometers covering 200~nm to 2500~nm. The whole system from integrating sphere entrance to total spectrum counts was calibrated for absolute spectral power density using NIST-traceable deuterium and tungsten lamps.

\begin{figure}[th]
\centering
\includegraphics{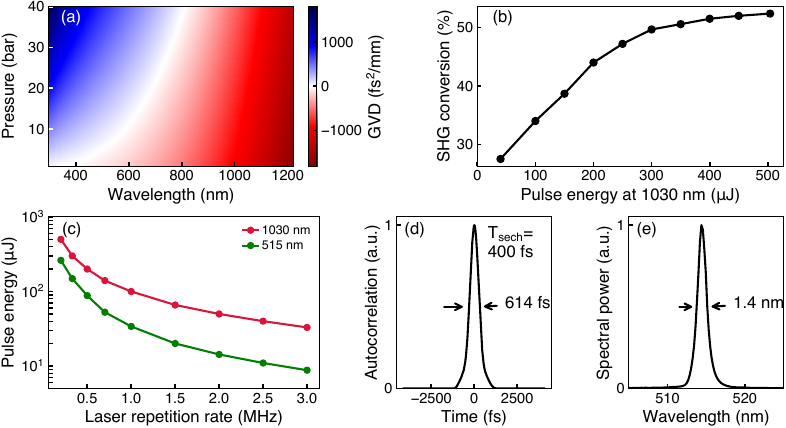}
\caption{(a) Calculated dispersion of a \SI{33}{\micro\meter} core diameter fiber filled with hydrogen or deuterium using the Marcatili model~\cite{marcatili_hollow_1964}. At moderate gas pressures ($<$40~bar), shorter wavelengths (e.g., 515~nm) lie within the normal dispersion regime (blue), while longer wavelengths (e.g., 1030~nm) lie in the anomalous regime (red). (b) SHG conversion efficiency as a function of input energy at 1030~nm. (c) Maximum pulse energy available at 1030~nm (red) and 515~nm (green) as a function of laser repetition rate. (d) Autocorrelation trace and (e) spectrum of the 515~nm pulses. The deconvolved pulse duration was 400~fs assuming a sech$^2$ temporal profile.}
\label{fig:SHG}
\end{figure}

As noted in Section~\ref{sec:intro}, the CtC process requires pumping in the normal dispersion regime. Figure~\ref{fig:SHG}(a) shows the calculated dispersion landscape for a \SI{33}{\micro\meter} core diameter fiber filled with hydrogen or deuterium, computed using the Marcatili model~\cite{marcatili_hollow_1964} for the fiber contribution and the model of Ref.~\cite{peck_refractivity_1977} for the gas contribution. At moderate pressures ($<$40~bar), the 515~nm pump wavelength lies in the normal dispersion regime (blue region), while the 1030~nm wavelength lies in the anomalous regime (red region). For 1030~nm pumping, normal dispersion would require pressures exceeding 110~bar. In contrast, pumping at 515~nm achieves normal dispersion at pressures above just 6~bar. The Marcatili model does not capture the dispersion perturbation introduced by the antiresonant cladding near its $m=1$ resonance, which for the wall thicknesses used here (160--174~nm) lies near 348--376~nm; the 515~nm pump is sufficiently far from this resonance that the normal-dispersion conclusion at the pump wavelength is unaffected.

The SHG conversion efficiency is shown in Fig.~\ref{fig:SHG}(b), reaching 55\% at maximum input energy and yielding up to \SI{275}{\micro\joule} at 515~nm. To prevent damage at the fiber input facet~\cite{lekosiotis_-target_2023}, the 515~nm pulse energy was limited to \SI{10}{\micro\joule} during experiments. Figure~\ref{fig:SHG}(c) shows the maximum available pulse energy at both 1030~nm (red) and 515~nm (green) as a function of repetition rate. The laser's pulse-picker architecture results in decreasing pulse energy at higher repetition rates, with the 515~nm energy falling to \SI{8}{\micro\joule} at 3~MHz. Consequently, the maximum repetition rate providing sufficient 515~nm pulse energy ($>$\SI{10}{\micro\joule}) was 2.5~MHz. Higher repetition rates could be accessed using a focused beam geometry in the SHG crystal, which would improve conversion efficiency at lower pulse energies. The 515~nm pulses were characterized by autocorrelation, as shown in Fig.~\ref{fig:SHG}(d). The measured full-width at half-maximum (FWHM) duration of 614~fs corresponds to a pulse duration of 400~fs assuming a sech$^2$ temporal profile. The spectral width was 1.4~nm FWHM [Fig.~\ref{fig:SHG}(e)].

\section{Single-pumping at 515~nm}\label{sec:SP}

\begin{figure}[ht]
\centering
\includegraphics{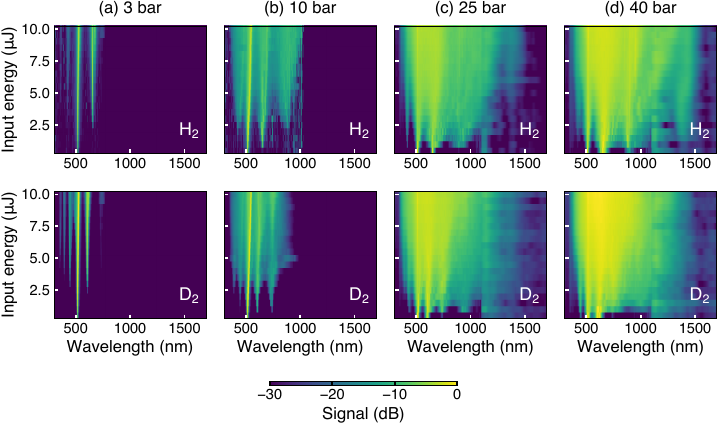}
\caption{Experimental output spectra on a logarithmic scale as a function of input energy at 515~nm for hydrogen (H$_2$, top) and deuterium (D$_2$, bottom) at increasing pressures. (a) At low pressure (3~bar), distinct Raman frequency combs are formed. (b) With increasing pressure (10~bar), the comb lines broaden and begin to merge. (c, d) At higher pressures (25 and 40~bar), the lines blend into broadband supercontinua.}
\label{fig:escans}
\end{figure}

The thin-wall fiber was initially pumped using only the 515~nm pulses, with hydrogen or deuterium as the filling gas. The laser repetition rate was set at 1~kHz. For increasing 515~nm energy up to \SI{10}{\micro\joule} and gas pressure up to 40~bar, the output spectra were recorded and calibrated, and are shown in Fig.~\ref{fig:escans}. The evolution of the output spectra with increasing pressure showcases the supercontinuum generation mechanism. A frequency comb is initially formed [see Fig.~\ref{fig:escans}(a)] through the vibrational Raman response. With increasing pressure the comb extends further, and at the same time individual lines broaden through self-phase modulation and the non-instantaneous rotational Raman response, eventually merging into a continuous spectrum [see Fig.~\ref{fig:escans}(b)]. Further increase of the gas pressure enables additional spectral extension into the infrared [see Fig.~\ref{fig:escans}(c)], until the uniform blending of all generated comb lines forms an ultra-flat supercontinuum at 40~bar of deuterium or hydrogen [see Fig.~\ref{fig:escans}(d)]. The sharp feature near 1100~nm visible in all measured spectra is an artifact of the junction between the two spectrometers used to record the output and is not a genuine spectral feature.

\begin{figure}[ht]
\centering
\includegraphics{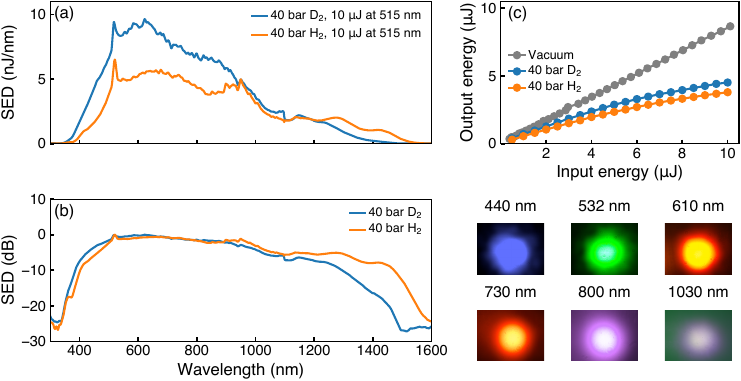}
\caption{(a) Supercontinuum generation at 40~bar hydrogen (H$_2$) or deuterium (D$_2$) using \SI{10}{\micro\joule} input energy at 515~nm and 1~kHz laser repetition rate plotted on a linear spectral energy density scale. (b) Same as (a) but on a logarithmic scale. (c) Energy transmission in vacuum (gray), 40 bar D$_2$ (green) and 40 bar H$_2$ (blue). Bottom right: far-field mode images obtained using band-pass filters at the annotated central wavelengths.}
\label{fig:515nmpumping}
\end{figure}

The generated supercontinua obtained at 40~bar of deuterium or hydrogen using \SI{10}{\micro\joule} of input energy at 515~nm are shown on a linear energy density scale in Fig.~\ref{fig:515nmpumping}(a).  The output spectra exhibit notable pump depletion and excellent flatness, even on a linear scale. Figure~\ref{fig:515nmpumping}(b) shows the same spectra on a logarithmic scale. The energy transmission was approximately 50\% in D$_2$ and 40\% in H$_2$ at \SI{10}{\micro\joule} input energy, as shown in Fig.~\ref{fig:515nmpumping}(c). Energy losses were due to linear attenuation in the fiber and nonlinear absorption in the gases. Coupling losses were measured at 10\% with the fiber evacuated, after correcting for Fresnel reflections at the windows [see gray dots in Fig.~\ref{fig:515nmpumping}(c)]. The modal quality of the generated supercontinuum was assessed across different bands using a microscope equipped with a camera and band-pass filters. The recorded far-field images shown in the bottom row of Fig.~\ref{fig:515nmpumping} are qualitatively consistent with predominantly fundamental-mode content at all measured wavelengths, with the 440~nm band showing some distortion due to its proximity to the resonance band; a quantitative modal decomposition or $M^2$ measurement is left to future work.

\begin{figure}[ht]
\centering
\includegraphics{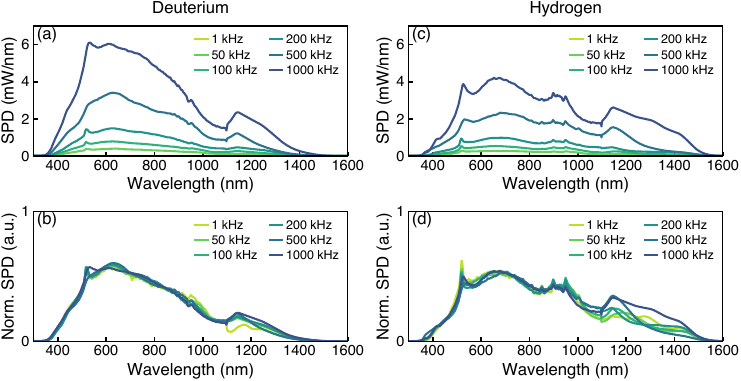}
\caption{Power scaling of the supercontinuum to 10~W input average power at 515~nm by incrementally increasing the laser repetition rate from 1~kHz to 1~MHz using \SI{10}{\micro\joule} input pulse energy at 515~nm and 40~bar of deuterium (a, b) or hydrogen (c, d). Output spectra are plotted on a linear SPD~(a,~c) and on a normalized SPD scale~(b,~d).}
\label{fig:SPSC_PS}
\end{figure}

The results shown above were generated at 1~kHz laser repetition rate, corresponding to an input average power at 515~nm of only 10~mW and an output spectral power density (SPD) on the order of a few~\SI{}{\micro\watt}/nm. To achieve high-power supercontinuum generation, the laser repetition rate was incrementally scaled to 1~MHz, corresponding to an input average power of 10~W at 515~nm. Using a fixed 515~nm input energy of \SI{10}{\micro\joule} and 40~bar of D$_2$, the output spectra recorded for increasing laser repetition rates are presented on a linear SPD scale in Fig.~\ref{fig:SPSC_PS}(a) and on a normalized SPD scale in Fig.~\ref{fig:SPSC_PS}(b) to observe relative shape differences. The corresponding output spectra using H$_2$ instead of D$_2$ are shown in Fig.~\ref{fig:SPSC_PS}(c, d). In both gases, the output power density was successfully scaled by three orders of magnitude to a few~mW/nm and the high-power output spectra closely resembled those obtained at lower repetition rates, with slight enhancement of the infrared side observed primarily in the H$_2$ spectra. Importantly, the power scaling showed no sign of saturation, raising the prospect of further scaling of power and repetition rate.

Overall, this Raman-driven CtC mechanism produced a broadband supercontinuum source with high spectral density in the visible and near-infrared, with a \SI{-10}{\decibel} span of 403--1469~nm in H$_2$ and 385--1295~nm in D$_2$, and access to the near-ultraviolet (down to $\sim$350~nm at the \SI{-20}{\decibel} level). The generated spectra exhibit excellent spectral flatness (due to strong pump depletion), high spatial quality, and scalability of the repetition rate (and thus input average power) to the MHz (watt) level.

\section{Dual-pumping at 515~nm and 1030~nm}\label{sec:DP}

Aiming to further enhance the infrared side of the supercontinuum, we tested dual-pumping the fiber with pulses at both 515~nm and 1030~nm. We emphasize that the two pumps drive qualitatively different nonlinear dynamics: at 40~bar the 515~nm pulse is in the normal dispersion regime and drives the CtC process described in Section~\ref{sec:SP}, whereas the 1030~nm pulse is in the anomalous dispersion regime [Fig.~\ref{fig:SHG}(a)] and undergoes soliton-like dynamics with pronounced Raman-induced red-shifting. The dual-pumped output is therefore best understood as the combination of CtC at 515~nm with separate Raman-soliton dynamics seeded by the 1030~nm pulse, together with cross-phase modulation and four-wave mixing between the two pumps. A detailed numerical analysis of the dual-pumped dynamics is beyond the scope of this work and is left as an avenue for future investigation. In the experimental implementation, the full setup of Fig.~\ref{fig:setup} was used to provide an additional pump arm at 1030~nm, which was spatio-temporally overlapped with the 515~nm pump at the input facet of the fiber. The thin-wall fiber was replaced with the thick-wall fiber (see Section~\ref{sec:setup}), shifting the high-loss resonance band from $\sim$348 to $\sim$376~nm and thus restricting the short-wavelength side of the supercontinuum in exchange for improved transmission in the infrared. Due to the absence of a dedicated focusing system, the coupling efficiency of the 1030~nm beam was limited to 65\%, in contrast to the 515~nm beam which was coupled with nearly 90\% efficiency in vacuum, as shown in Fig.~\ref{fig:515nmpumping}(c).

\begin{figure}[ht]
\centering
\includegraphics{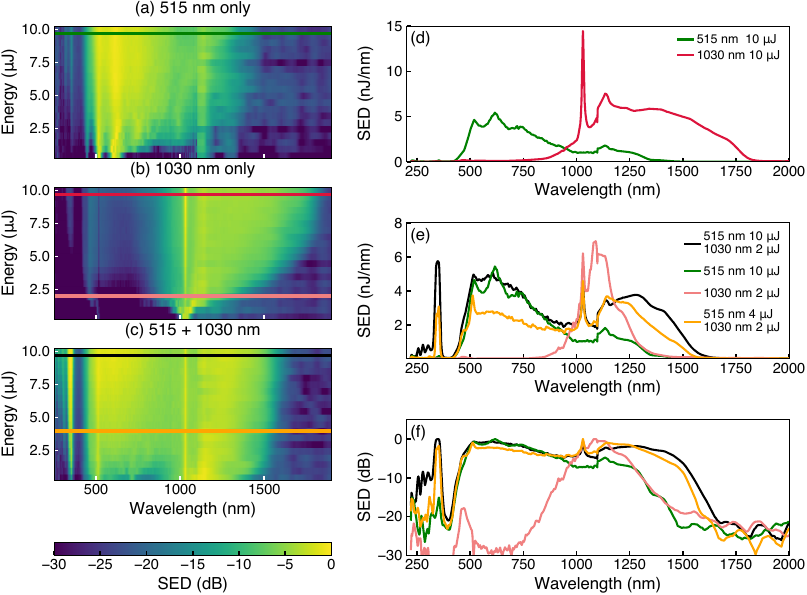}
\caption{Spectral evolution with increasing input energy at 40~bar D$_2$ using the thick-wall fiber (high-loss resonance band at $\sim$376~nm). (a) Single-pumping at 515~nm with increasing energy up to \SI{10}{\micro\joule}. (b) Single-pumping at 1030~nm with increasing energy up to \SI{10}{\micro\joule}. (c) Dual-pumping with increasing 515~nm energy up to \SI{10}{\micro\joule} and fixed 1030~nm energy at \SI{2}{\micro\joule}. (d) Selected spectra from (a) and (b) plotted on a linear scale (line colors match the horizontal bars in (a) and (b)). (e) Selected spectra from (a), (b), and (c) plotted on a linear scale (line colors match the horizontal bars in (a), (b) and (c)). (f) Same as (e) but on a logarithmic scale.}
\label{fig:DP}
\end{figure}

Operating at 40~bar D$_2$ and 1~kHz laser repetition rate, the performance of the thick-wall fiber was initially tested using the 515~nm pump only, as shown in Fig.~\ref{fig:DP}(a). At maximum energy (\SI{10}{\micro\joule}), the generated supercontinuum [shown as a green line in Fig.~\ref{fig:DP}(d)] was similar to that recorded using the thin-wall fiber [green line in Fig.~\ref{fig:515nmpumping}(a)], except for the lower spectral energy density in the visible region (from 10 to \SI{5}{\nano\joule\per\nano\meter}) and the shift of the short-wavelength limit from 350 to 400~nm, confirming the effect of this fiber's increased wall thickness. 

Blocking the 515~nm beam, we launched the 1030~nm beam alone to record the spectral evolution at 40~bar D$_2$, as shown in Fig.~\ref{fig:DP}(b). Here, the 1030~nm pump lay in the anomalous dispersion regime [see Fig.~\ref{fig:SHG}(a)] and the spectral broadening exhibited strong red-shifting with increasing energy, up to 1800~nm. Although a small fraction of the energy was also transferred toward the visible region, with a noticeable feature at 480~nm, spectral broadening was most prominent in the infrared, with a sharp undepleted peak at 1030~nm as shown on a linear scale in red in Fig.~\ref{fig:DP}(d). 

The combination of the two pumps was then tested, with the evolution of the output spectra shown in Fig.~\ref{fig:DP}(c). Here, the 515~nm energy was varied up to \SI{10}{\micro\joule} while the 1030~nm energy was fixed at \SI{2}{\micro\joule}. Although single-pumping with \SI{2}{\micro\joule} at 1030~nm generated a spectrum with \SI{-10}{\decibel} bandwidth of 940--1308~nm [pink line in Fig.~\ref{fig:DP}(e) and (f)], when combined with 515~nm pulses it produced an ultra-flat broadband supercontinuum with \SI{-10}{\decibel} bandwidth spanning 439--1547~nm [black line in Fig.~\ref{fig:DP}(e) and (f)]. The optimal 1030~nm energy of \SI{2}{\micro\joule} was determined based on the flatness of the supercontinuum; higher energies produced sharper undepleted peaks at 1030~nm, similar to that shown in Fig.~\ref{fig:DP}(d).

The two main features of the dual-pumped supercontinuum compared to the single-pumped case were the significant enhancement of the 1000--1650~nm range and a strong resonance-enhanced four-wave mixing~\cite{Tani_2018} peak at 350~nm. Furthermore, at wavelengths below this peak there was a significant increase in the spectral energy density, down to 200~nm, most easily observable on a logarithmic scale [see Fig.~\ref{fig:DP}(f)]. The enhancement of the infrared portion of the spectrum is attributed to the Raman-soliton dynamics seeded by the 1030~nm pump, which extends the supercontinuum further into the infrared while also enhancing the spectral energy density in this region. The strong four-wave mixing peak at 350~nm is attributed to a resonance-enhanced process involving the 515~nm and 1030~nm pumps. The weaker deep ultraviolet continuum between 200~nm and 350~nm could also be due to four-wave-mixing conversion of the spectrally broadened pumps~\cite{belli_strong_2019, belli_broadband_2020}; however, the ripple structure suggests a residual anti-Stokes Raman frequency comb arising from the 350~nm peak~\cite{gao_raman_2022}.

Another notable feature of the dual-pumped configuration was the strong spectral broadening even at very low 515~nm input energies below \SI{2.5}{\micro\joule}, which established the edges of the supercontinuum [see Fig.~\ref{fig:DP}(c)]. High spectral flatness was achieved for pulse energies just above \SI{2.5}{\micro\joule}. The combination of lower 515~nm energies with 1030~nm pulses can significantly reduce the net energy requirements of the system while simultaneously enhancing the infrared side and preserving the visible portion of the supercontinuum. For instance, pumping with \SI{4}{\micro\joule} at 515~nm and \SI{2}{\micro\joule} at 1030~nm delivers an ultra-flat broadband supercontinuum [orange line in Fig.~\ref{fig:DP}(e)] using a total laser energy at 1030~nm of just \SI{10}{\micro\joule}, assuming 50\% second harmonic conversion efficiency. This is half the net laser energy required when single-pumping with \SI{10}{\micro\joule} at 515~nm [see green line in Fig.~\ref{fig:DP}(e)], without compromising supercontinuum performance.

\begin{figure}[ht]
\centering
\includegraphics{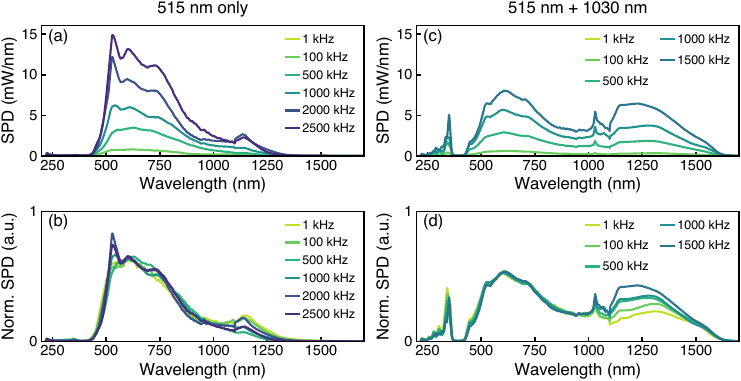}
\caption{Power scaling of the supercontinuum using the thick-wall fiber and 40~bar D$_2$. (a) Single-pumping at 515~nm with increasing repetition rate up to 2.5~MHz, plotted on a linear SPD scale. (b) Normalized SPD scale showing spectral shape evolution with increasing power. (c) Dual-pumping at 515~nm and 1030~nm with increasing repetition rate up to 1.5~MHz, plotted on a linear SPD scale. (d) Normalized SPD scale for dual-pumping.}
\label{fig:DP_PS}
\end{figure}

We tested power scaling of both the single-pumped (515~nm only) and dual-pumped supercontinua shown in Fig.~\ref{fig:DP}(e). Although power scaling of the 515~nm pumped supercontinuum was previously performed using the thin-wall fiber (see Fig.~\ref{fig:SPSC_PS}), those tests were conducted only up to 1~MHz laser repetition rate (10~W input average power at 515~nm), without pushing the source to its limits. As discussed in Section~\ref{sec:setup}, the available 515~nm pulse energy was sufficient to operate up to 2.5~MHz repetition rate (25~W input average power at 515~nm). These measurements were made with the thick-wall fiber ($\sim$174~nm wall thickness) used for the dual-pumping experiments, rather than the thin-wall fiber ($\sim$160~nm) used for Fig.~\ref{fig:SPSC_PS}, so Figs.~\ref{fig:SPSC_PS} and \ref{fig:DP_PS} are not directly comparable. The power scaling results, using 40~bar of deuterium, are shown in Fig.~\ref{fig:DP_PS}. Power scaling of the single-pumped 515~nm supercontinuum was successfully achieved up to 2.5~MHz, with spectral power density exceeding 5~mW/nm over 491--869~nm and 10~mW/nm over 509--769~nm [see Fig.~\ref{fig:DP_PS}(a)]. The integrated output average power at 2.5~MHz was approximately 5~W, corresponding to a fiber transmission of $\sim$20\% at this repetition rate (reduced from $\sim$35\% at 1~kHz). The spectral shape evolution is shown on a normalized power density scale in Fig.~\ref{fig:DP_PS}(b), where small variations near the pump and on the infrared side occur with increasing power. The dual-pumped supercontinuum was successfully scaled to 1.5~MHz [see Fig.~\ref{fig:DP_PS}(c)], delivering an integrated output average power of approximately 5~W with spectral power density exceeding 2.5~mW/nm over a 990~nm bandwidth (479--1469~nm)---matching the single-pumped 2.5~MHz output power at substantially lower 515~nm input. The output spectral shape remained nearly identical except for the infrared region, which was gradually enhanced with increasing power. As before, there was no sign of saturation.

\section{Conclusion}
We have demonstrated Raman-driven supercontinuum generation with high flatness and broad bandwidth in hydrogen- and deuterium-filled antiresonant fibers, using both single-pumping at 515~nm and dual-pumping at 515~nm and 1030~nm. The supercontinuum sources were successfully power-scaled to 2.5~MHz repetition rate and 25~W input average power, with a spectral power density exceeding 5~mW/nm over 491--869~nm and 10~mW/nm over 509--769~nm, and a total output average power of approximately 5~W. The dual-pumped configuration delivered comparable output average power ($\sim$5~W) at 1.5~MHz with substantially lower 515~nm input, exceeding 2.5~mW/nm over a 990~nm bandwidth (479--1469~nm). Notably, the spectral shape remained largely unchanged across three orders of magnitude in repetition rate, from 1~kHz to 2.5~MHz, indicating that inter-pulse thermal effects in hydrogen and deuterium are manageable at these parameters. Further extension of the short-wavelength edge was limited only by fiber resonances, and the far-field profiles recorded in different spectral bands are qualitatively consistent with predominantly fundamental-mode content across the bandwidth.

To our knowledge, these results represent the first scaling of the CtC mechanism to watt-level average powers. In contrast to methane-filled fibers, where thermal decomposition of the gas restricted operation to repetition rates below approximately 50~kHz~\cite{plosz_supercontinuum_2025}, no signs of gas degradation or fiber damage were observed here, even at 2.5~MHz. Importantly, no sign of saturation was observed in either configuration, indicating that neither the fiber nor the gas medium had reached a fundamental limit. The power scaling achieved here was constrained only by the available pulse energy from the laser at higher repetition rates. The far-field mode profiles recorded across the supercontinuum bandwidth confirm that the output retains high spatial quality.

Several avenues remain open for future work. Detailed numerical modeling of the CtC process, and in particular of the dual-pumped configuration, would help to disentangle the relative contributions of cross-phase modulation, four-wave mixing and independent soliton dynamics seeded by the 1030~nm pulse, and to identify whether instantaneous Kerr or non-instantaneous rotational Raman scattering dominates the line-filling step---a question which is in principle accessible experimentally in hydrogen and deuterium, where the rotational and vibrational responses are spectrally distinct. The shot-to-shot stability, relative intensity noise, and temporal coherence of the generated supercontinuum also warrant detailed investigation; importantly, although the present demonstration is seeded from quantum noise, this is due to the specific implementation rather than to the CtC mechanism itself, and we anticipate that variants of the process should be capable of producing a fully coherent supercontinuum, in contrast to modulational-instability-driven sources. Spatial mode quality (quantified via $M^2$ or a full modal decomposition) and polarization extinction ratio across the supercontinuum bandwidth, both of which directly affect application performance, represent natural next measurements for this source. The thermal and refractive-index perturbations that build up in the gas at high repetition rates~\cite{koehler_long-lived_2018, koehler_post-recombination_2021, tani_long-lived_2025} will eventually set an upper bound on the achievable average power; mapping out this boundary for the CtC process, in both single- and dual-pumped configurations, is an important next step. From an engineering perspective, the use of hydrogen at elevated pressures will necessitate robust sealed gas-cell designs for operation outside controlled laboratory settings.

Notwithstanding these challenges, the results presented here establish hydrogen- and deuterium-filled antiresonant fibers as a promising platform for high-power, ultra-broadband spatially coherent supercontinuum light sources.

\begin{backmatter}
\bmsection{Funding}
JCT is funded by a Chair in Emerging Technologies from the Royal Academy of Engineering and by the Institution of Engineering and Technology (IET) through the IET A F Harvey Engineering Research Prize. FB is supported by the Royal Academy of Engineering under the Research Fellowship grant No. RF/202021/20/310.

\bmsection{Disclosures}
The authors declare no conflicts of interest.

\bmsection{Data availability}
Data underlying the results presented in this paper are not publicly available at this time but may be obtained from the authors upon reasonable request.
\end{backmatter}

\bibliography{extracted_refs}

\end{document}